\documentclass[10pt,twocolumn,twoside]{IEEEtran}
\IEEEoverridecommandlockouts

\usepackage{cite}
\usepackage{amsmath,amssymb,amsfonts}
\usepackage{algorithmic}
\usepackage{graphicx}
\usepackage{textcomp}
\usepackage{balance}
\usepackage[usenames,dvipsnames]{xcolor}
\usepackage[utf8]{inputenc}
\usepackage{subcaption}

\usepackage{tabularx}
\usepackage{colortbl}
\definecolor{softgray}{gray}{0.8}

\def\BibTeX{{\rm B\kern-.05em{\sc i\kern-.025em b}\kern-.08em
    T\kern-.1667em\lower.7ex\hbox{E}\kern-.125emX}}
\usepackage{lipsum}

\title{\LARGE \bf 
Bifurcation Analysis of Sub-Synchronous Oscillations Related to Grid-Forming Converter Inner Controllers
}

\author{Luke Ian Benedetti$^\dagger$, Robin Preece$^\dagger$, Panagiotis N. Papadopoulos$^\dagger$
\vspace{-0.5cm}
 \thanks{ 
$^\dagger$All authors are with the Department of Electrical and Electronic Engineering, The University of Manchester, Manchester, UK. Emails: { \{luke.benedetti; robin.preece; panagiotis.papadopoulos\}@manchester.ac.uk}}
\thanks{Financial support is acknowledged from a UKRI Future Leaders Fellowship MR/Y00390X/1 (P. N. Papadopoulos and L. I. Benedetti) and EPSRC-funded Supergen Energy Networks Hub Project EP/S00078X/2 (R. Preece). All results can be fully reproduced using the methods and data described in this paper and provided references.
For the purpose of open access, the authors have applied for a Creative Commons Attribution (CC BY) license to any Author Accepted Manuscript version arising from this submission.}
}

\graphicspath{{Figures/}}

\begin{document}
\begingroup
\allowdisplaybreaks

\maketitle

\begin{abstract}
To ensure power system stability and security, it is vital to understand the complex nonlinear power system dynamics related to converter-interfaced generators.   
For example, grid-forming (GFM) converters are expected to be a key asset for maintaining a strong and stable power system, but might cause wide-bandwidth stability issues with underlying mechanisms heretofore unseen or understudied, including sub-synchronous oscillations (SSOs). 
This paper details a continuation-based bifurcation analysis of a GFM converter, 
revealing stability bounds with respect to operational conditions in addition to the time constant of the cascaded inner voltage and current controllers. 
We focus our analysis on the strong grid instability caused by an inner controller-related SSO, including continuation of the limit cycle past the Hopf bifurcation point, revealing rapid onset of unacceptably large oscillations. 
Furthermore, we investigate the impact of the circular current limiter, revealing spurious Hopf bifurcations in weak grids associated with the aforementioned SSO when adopting smooth approximations; this suggests the need for careful implementation of such approximations for GFMs, at least in bifurcation studies. 
\end{abstract}


\section{Introduction}

Grid-forming (GFM) devices are required for the formation and regulation of the voltage-sourced power system \cite{Li2022}. As such, GFM converters are expected to play a major role in the energy generation mixture \cite{Badrzadeh2024}
as fossil-fuelled synchronous generators are increasingly disconnected to meet critical decarbonisation goals \cite{IEA2024}. Such technologies bring novel, wideband dynamics spanning large frequency ranges \cite{Shair2021} whose characteristics need to be fully understood to ensure desirable dynamic responses, avoid adverse interactions, oscillations, or instability, and design methods for the monitoring and mitigation thereof. 
Furthermore, single operating point small-signal analyses rely on localised linear assumptions, meaning they cannot capture important nonlinear dynamical phenomena including bifurcations and limit cycles \cite{Strogatz2018, Kuznetsov1998, Datseris2022}. 
As such, this paper performs a continuation-based bifurcation analysis of a standard GFM converter with cascaded control structure, revealing the nonlinear effects of operating conditions and inner controller time constants. 

The standard cascaded GFM control structure \cite{Baeckeland2025} has known limitations with respect to inner controller tuning and strong grid instabilities \cite{Sharjeel2025}. In particular, the control system incites sub-synchronous oscillations (SSOs) related to the cascaded inner voltage and current controllers, seen to be around $12.5~\text{Hz}$ in \cite{Sharjeel2025_2}, and $15~\text{Hz}$ to $17~\text{Hz}$ in \cite{Benedetti2025}.
Using a simplified model which allowed for analytical expression of the damping ratio of the SSO, the authors of \cite{Sharjeel2025_2} reveal the potential destabilising effect of increasing the current controller time constant or the filter capacitance, or decreasing the inner voltage controller proportional gain. 
Under different parametric conditions but with similar controller architecture, \cite{Hong2023} finds slower strong grid instability-inducing SSOs of around $2.5~\text{Hz}$ which are related to interactions between the inner voltage and the outer synchronisation controllers. Meanwhile, \cite{Li2024} suggests increasing the voltage controller bandwidth to mitigate such issues. Whereas, in this paper we focus on the cascaded inner controller-related SSO, and investigate for the first time the parametric stability bounds of the (full detail) GFM converter with respect to codimension (codim)-2 continuation of the strong grid Hopf point of the SSO considering the time constants (bandwidths) of the inner voltage and current controllers simultaneously. We also consider stability bounds across a range of operational parameter variations with respect to the short-circuit ratio (SCR), X/R ratio, and active power set-point of the GFM.

Going beyond local equilibrium point bifurcation analysis, continuation of the limit cycle of a Hopf bifurcation has been seen to offer insights into the large-disturbance stability of a converter-based system \cite{Ma2023,Zhao2025}.
For example, we might observe a small region of attraction (RoA),
caused by the unstable limit cycle that converges towards the stable equilibrium point prior to a subcritical Hopf bifurcation \cite{Ma2023,Zhao2025}. Additionally, following a supercritical Hopf bifurcation, a stable limit cycle appears around the unstable equilibrium point \cite{Ma2023,Zazo2020}. If said limit cycle is very small in magnitude then the system states (or observables) may stay well within the acceptable bounds, meaning by the formal power system stability definition \cite{Hatziargyrou2021} the system is stable, despite the local equilibrium point analysis suggesting instability. Both of these scenarios are ruled out for the GFM system in this work as we use normal form theory and limit cycle continuation to reveal that the strong grid Hopf point is supercritical with the emerging limit cycle quickly increasing to unacceptable levels.

Since continuation analysis typically requires a continuously differentiable vector field \cite{Kuznetsov1998, Datseris2022}, non-smooth elements such as saturation limiters require smooth approximations to be included in the analysis \cite{Moutevelis2022}. 
Therefore, we investigate the influence of including corresponding smooth approximations on the results of our bifurcation analyses. We find that smooth approximations of the circular current limiter can cause spurious Hopf bifurcations of the aforementioned SSO. 
This can invalidate the resultant bifurcation studies.


Ultimately, we add to the growing literature pertaining to nonlinear dynamics, bifurcations and SSOs of GFM converters with a systematic, continuation-based bifurcation analysis of the standard cascaded control structure and concomitant SSOs. The contributions of this work include:
\begin{itemize}
    \item Revealing stability bounds of GFM converters for key operational and control parameters via local bifurcation analysis. In particular, to the best of our knowledge, this has not previously been performed for the simultaneous tuning of the inner controller time constants.
    \item Unveiling the strong grid Hopf bifurcation of the inner controller-related SSO of the GFM to be supercritical with rapid onset of an unacceptably large limit cycle. 
    \item Investigation of the impact of the circular current limiter smooth approximation, revealing potential for spurious Hopf bifurcations of the aforementioned SSO.
    \item We provide the Julia scripts for: modelling the GFM converter with ModelingToolkit.jl \cite{ma2021modelingtoolkit}; continuation analysis with BifurcationKit.jl \cite{veltz2020}, including bespoke wrappers required to enable this functionality; and nonlinear dynamical simulations with DifferentialEquations.jl \cite{rackauckas2017differentialequations}.
\end{itemize}
  


\section{Methodology}
\label{sec:meth}
The analysis methodology adopted in this work is based on local (equilibrium point) bifurcation analysis \cite{Strogatz2018}. In particular, we utilise a standard predictor-corrector continuation approach \cite{Keller1988,veltz2020}. In addition to this, continuation of the limit cycle is performed based on the collocation method \cite{Datseris2022, veltz2020}.

In this work, we consider a nonlinear dynamical system representation in ordinary differential equation (ODE) form,
\begin{equation}
    \dot{\boldsymbol{x}} = \boldsymbol{f}(\boldsymbol{x},\boldsymbol{p}),
\end{equation}
\noindent where $\boldsymbol{f}:\mathbb{R}^N \times \mathbb{R}^{N_p} \rightarrow \mathbb{R}^N$ denotes a sufficiently smooth vector field \cite{Kuznetsov1998} with $N$ dynamic states, $\boldsymbol{x}$, and $N_p$ parameters, $\boldsymbol{p}$. The dot operator signifies differentiation w.r.t. time.

\subsection{Bifurcation Theory}
\label{sec:meth:bifur}
Local bifurcation theory deals with the continuous change of parameters, finding points at which the qualitative behaviour of the nonlinear dynamical system suddenly changes apropos the stability of the equilibrium point \cite{Kuznetsov1998,Chen2023} evaluated at
\begin{equation}
\label{eq:init}
    0 = \boldsymbol{f}(\boldsymbol{x},\boldsymbol{p}).
\end{equation}

Such a change is reflected in the eigenvalues of the Jacobian of $\boldsymbol{f}(\boldsymbol{x})$ (i.e., evaluated at $\boldsymbol{p}$), often termed the state matrix $\boldsymbol{A}$, crossing into the right half-plane. With respect to codim-1 parameter variations \cite{Kuznetsov1998}, a single real-valued eigenvalue crossing results in aperiodic instability \cite{Kundur} and might indicate a pitchfork, transcritical, or saddle node bifurcation. However, the latter of these is more typical in power system scenarios in the form of voltage instability (e.g., maximum loadability limits), which from the nonlinear dynamical analysis perspective represents the point at which a stable and unstable equilibrium point collide and subsequently vanish \cite{Kuznetsov1998}\cite{CutsemVournas}. 

A pair of complex conjugate eigenvalues of $\boldsymbol{A}$ crossing the imaginary axis indicates a Hopf bifurcation, leading to oscillatory instability. A non-degenerate Hopf bifurcation \cite{Kuznetsov1998} can be either supercritical or subcritical. For the former, a stable limit cycle emerges around the (now unstable) equilibrium point. The latter means an unstable limit cycle has collapsed onto the stable equilibrium point and the limit cycle disappears while the equilibrium point becomes oscillatory unstable (but without a nearby stable limit cycle as in the supercritical case). 

Although not encompassed within the theory of bifurcations of continuously differentiable ODEs, the limit-induced bifurcation is an important consideration in modern power systems \cite{Xing2021}. When a saturation limiter engages, the system undergoes a non-smooth change of its vector field. This essentially blocks the controllers ``behind'' the saturation limiter thereby eliminating the corresponding equations and dynamic states. \cite{Xing2021} highlights that such a change of the governing equations
can immediately shift the system to an unstable state.
However, non-smooth action is not compatible with standard continuation methods \cite{Kuznetsov1998}, thereby incentivising the use of smooth approximations for non-smooth elements \cite{Moutevelis2022}. The smooth approximation used in this work is discussed in more detail in Section \ref{sec:mod:limit}. 

\subsection{Continuation of Equilibrium Branch or Bifurcation Branch}
\label{sec:meth:cont}
Starting from a known initial condition, continuation methods follow the equilibrium branch, along which (\ref{eq:init}) holds true, as a parameter is varied. Fundamentally, this involves a predictor and a corrector step \cite{Datseris2022}, with one of the most common and robust algorithms being pseudo-arclength continuation (PALC) as implemented in this work using BifurcationKit.jl (BFK) \cite{veltz2020}. Bifurcation points are identified based on the behaviour of the eigenvalues of the Jacobian as discussed in Section \ref{sec:meth:bifur}.

Once a bifurcation point is identified, a second continuation parameter can be selected and they can be varied simultaneously along the bifurcation curve for which the relevant eigenvalue(s) remain on the imaginary axis. If no other bifurcation branches are found within the (codim-2) parameter space considered, this bifurcation branch represents the stability boundary. This procedure is performed for the strong grid Hopf bifurcation in this work.



\subsection{Hopf Normal Form and Limit Cycle Continuation}
\label{sec:meth:contPO}

In this work, we find that the strong grid Hopf bifurcation is supercritical, meaning that a limit cycle emerges after the Hopf point. This is determined by considering the centre manifold theorem \cite{Kuznetsov1998} which states that the dynamics of the system can be reduced to a two-dimensional invariant manifold\footnote{This is the centre manifold. It can be noted that, at the bifurcation point, it is tangent to the plane spanned by the corresponding eigenvectors \cite{Kuznetsov1998}.} near the Hopf bifurcation point. The normal form of the Hopf bifurcation can be calculated by considering the third-order Taylor series approximation of the centre manifold, which captures the nonlinear effects that determine whether the Hopf bifurcation is subcritical, supercritical, or degenerate (Section \ref{sec:meth:bifur}). In particular, the Hopf normal form  as considered in BFK is
\begin{equation}
    \label{eq:HopfNormalForm} \dot{\boldsymbol{z}}=\boldsymbol{z}\left(j\omega+a\,\delta p+b\left|\boldsymbol{z}\right|^2\right),
\end{equation}
\noindent where $\boldsymbol{z}$ is a complex phasor, $\omega$ is the frequency of the complex conjugate eigenvalue pair giving rise to the Hopf bifurcation, $a$ quantifies the linear growth of the limit cycle, which holds true for very small values of the change in continuation parameter \cite{Kuznetsov1998}, $\delta p$. The first Lyapunov coefficient is given by $b$, whose real part indicates supercriticality if negative, subcriticality if positive, and degeneracy if zero.  

We can then apply continuation to the limit cycle itself\footnote{A detailed technical discussion of this is out of the scope of this paper and the reader is referred to \cite{veltz2020,Kuznetsov1998} for more information.} as the parameter continues to change. This requires an estimation of the limit cycle, as represented as a boundary value problem, achieved in this case using the orthogonal collocation method \cite{Doedel2005X}, again implemented with BFK \cite{veltz2020}. Therefore, as we further change the continuation parameter, we can track the period and envelope of the limit cycle. In our (supercritical) case, this is done to check whether the resulting oscillation remains small or quickly increases to unacceptably large values (e.g., extensive voltage and current swings).

\section{Grid-Forming Converter System Under Test}
\label{sec:mod}
All models were implemented in the Julia programming language using ModelingToolkit.jl (MTK) within the SciML ecosystem\footnote{The corresponding modelling (and analysis) scripts are available in \cite{benedetti_2026_data}. Note, bespoke wrappers for extraction of the vector field and Jacobians of a hierarchically constructed MTK model are provided to allow for continuation analysis with the BifurcationKit.jl package \cite{veltz2020}.} \cite{ma2021modelingtoolkit}.
The nominal parameters are displayed in Table \ref{tab:params} and described further throughout this section. The $^*$ superscript denotes references and set points.


\renewcommand{\arraystretch}{1.1}
\begin{table}[hbtp]
    \centering
    \caption{\centering \fontsize{8pt}{8pt}\selectfont \textsc{Nominal Parameters of the GFM-Infinite Bus System}}
    \begin{tabularx}{\columnwidth}{>{\centering\arraybackslash}X >{\centering\arraybackslash}X
    | >{\centering\arraybackslash}X
     >{\centering\arraybackslash}X
    }
        \hline
       \textbf{Parameter} & \textbf{Value}  & \textbf{Parameter} & \textbf{Value} \\
       \hline
       \arrayrulecolor{softgray}
       $P^*$ & $0.8~\text{pu}$ & $m_p$ & $0.01~\text{pu}$ \\
       \hline
       $Q^*$ & $0~\text{pu}$ & $\omega_{c,p}$ & $2\pi5~\text{rad\,s}^{-1}$ \\
       \hline
       $V^*$ & $1~\text{pu}$ & $m_q$ & $0.01~\text{pu}$ \\
       \hline
       $\omega_b$ & $2\pi50~\text{rad\,s}^{-1}$ & $\omega_{c,q}$ & $2\pi5~\text{rad\,s}^{-1}$ \\
       \hline
       $R_f$ & $0.03~\text{pu}$ & $\zeta_{IVC}$ & $0.7$ \\
       \hline
       $X_{l,f}$ & $0.08~\text{pu}$ & $\tau_{IVC}$ & $45~\text{ms}$ \\
       \hline
       $X_{c,f}$ & $13.51~\text{pu}$ & $\zeta_{ICC}$ & $0.7$ \\
       \hline
       $SCR$ & $2.5$ & $\tau_{ICC}$ & $1.5~\text{ms}$ \\
       \hline
       $X/R$ & $10$ & $|\boldsymbol{i}|_{\max}$ & $1.2~\text{pu}$ \\
       \hline
       $~$ & $~$ & $T_s$ & $47.12~\text{ms}$ \\
       \hline
    \end{tabularx}
    \label{tab:params}
\end{table}

The models of the electrical and GFM control systems are completed in independent rotating reference frames, synchronised to the grid voltage angle, $\angle \boldsymbol{v_g}$, and virtual rotor speed, $\theta$, respectively. A phasor can then be denoted in the form ${\boldsymbol{\nu}=\nu_d+j\nu_q}$ and physical signals in the control frame are discerned by superscript $^{ctrl}$. Conversion of signals between reference frames is through a geometric transform \cite{Moutevelis2022}.

\subsection{Electrical System}
We consider a GFM voltage-sourced converter (VSC) connected to a Th\'{e}venin equivalent grid model. The single line circuit diagram is illustrated in Fig. \ref{fig:GFMIB},
\begin{figure}
    \centering
    \includegraphics[width=0.75\columnwidth,trim=145 715 140 50, clip]{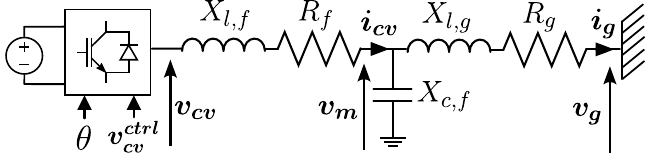}
    \caption{Single line circuit diagram of grid-forming converter connected to Th\'{e}venin equivalent grid model.}
    \label{fig:GFMIB}
\end{figure}
where $\boldsymbol{v_{cv}}$, $\boldsymbol{v_{m}}$, and $\boldsymbol{v_{g}}$ are the voltages at the converter switches, across the filter capacitor, and of the Th\'{e}venin equivalent infinite bus, respectively. The current through the coupling filter and grid impedance are denoted by $\boldsymbol{i_{cv}}$ and $\boldsymbol{i_{g}}$, respectively. The VSC is the averaged model with $\theta$ and $\boldsymbol{v_{cv}^{ctrl}}$ denoting the virtual rotor angle of the GFM and the requested value of $\boldsymbol{v_{cv}}$ in the controller reference frame, respectively. Furthermore, we consider that, on the timescales of interest, there is sufficient power available from the primary source to compensate for any current drawn, thereby allowing representation with an ideal voltage source on the DC side of the converter.

The per unit resistance, inductive reactance, and capacitive reactance of the coupling filter are denoted by $R_f$, $X_{l,f}$, and $X_{c,f}$, respectively. The resistance and inductive reactance of the Th\'{e}venin equivalent grid are $R_g$ and $X_{l,g}$ but when it comes to continuation, we are more interested in these parameters from the perspective of the short circuit ratio ($SCR$) and the $X/R$ ratio, with the corresponding conversion
\begin{align}
\label{eq:SCR}
    SCR &= 1/\sqrt{\left(R_g^2+X_{l,g}^2\right)},\\
     X/R &= X_{l,g} / R_g. 
\end{align}

\vspace{-5mm}
\subsection{Droop Control}
The full GFM control system block diagram is illustrated in Fig. \ref{fig:GFMcontrol}.
\begin{figure}
    \centering
    \includegraphics[width=0.85\columnwidth,trim=25 637 265 17, clip]{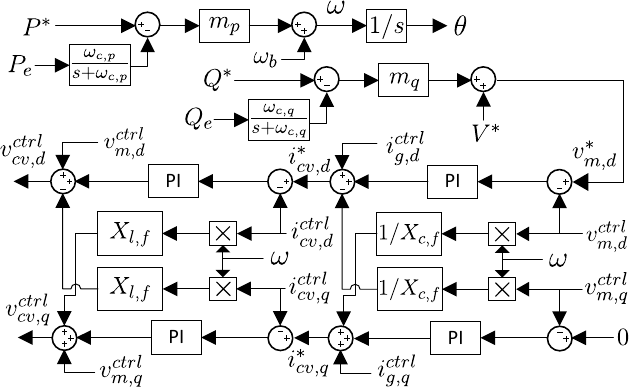}
    \caption{Multi-loop, droop-based GFM controller.}
    \label{fig:GFMcontrol}
\end{figure}
The active power/synchronisation controller is a standard droop with low-pass filter. The droop gain and cut-off frequency are denoted by $m_p$ and $\omega_{c,p}$, respectively. Also included is a reactive power droop which augments the voltage magnitude reference with droop gain and cut-off frequency denoted by $m_q$ and $\omega_{c,q}$, respectively.

\subsection{Inner Voltage and Current Controllers}
\label{sec:mod:inner}
The inner voltage and current controllers adopt a vector control architecture including proportional-integral (PI) regulators in addition to feed-forward and decoupling terms \cite{Baeckeland2025}. The proportional and integral gains are denoted by $K_{p,IVC}$ and $K_{i,IVC}$, and $K_{p,ICC}$ and $K_{i,ICC}$ for the inner voltage and current PI controllers, respectively.
Tuning of the inner controllers is achieved independently using pole-placement.

For the inner current controller, assuming ideal timescale separation, neglection of pulse-width modulation (PWM)/control delay, and considering identical d-axis and q-axis control loops \cite{Yazdani}, the closed loop consists of the transfer functions of the PI controller, 
\begin{equation}
    G_{PI}(s) = K_{p,ICC} + \frac{K_{i,ICC}}{s},
\end{equation} 
\noindent and the RL branch of the output filter, 
\begin{equation}
    G_{RL}(s) = \frac{1}{(X_{l,f}/\omega_b)s+R_f},
\end{equation} 
\noindent with unity feedback of the measured output current, $\boldsymbol{i_{cv}}$. The resulting characteristic equation is
\begin{equation}
    s^2+\frac{R_f+K_{p,ICC}}{(X_{l,f}/\omega_b)}s+\frac{K_{i,ICC}}{(X_{l,f}/\omega_b)}=0.
\end{equation}
\noindent We can then compare this to the standard second-order characteristic equation with natural frequency of $\omega_{n,ICC}$ and damping ratio of $\zeta_{ICC}$, giving PI controller gains in the form
\begin{align}
\label{eq:kpicc}
    K_{p,ICC} &= 2\,\zeta_{ICC}\,\omega_{n,ICC}\,\frac{X_{l,f}}{\omega_b}-R_f, \\
    \label{eq:kiicc}
    K_{i,ICC} &= \omega_{n,ICC}^2\frac{X_{l,f}}{\omega_b}.
\end{align}
\noindent The speed of the closed-loop response can then be chosen by considering the relation between $\omega_{n,ICC}$, $\zeta_{ICC}$ and the $5\%$ settling time, termed $\tau_{ICC}$ for the inner current controller, based on the approximate relation
\begin{equation}
\label{eq:tauicc}
    \tau_{ICC} \approx \frac{3}{\zeta_{ICC}\,\omega_{n,ICC}}.
\end{equation}
\noindent Rearranging (\ref{eq:tauicc}) for $\omega_{n,ICC}$ and substituting back into (\ref{eq:kpicc}) and (\ref{eq:kiicc}) allows us to tune our closed-loop response based only on $\zeta_{ICC}$ (which we keep constant at 0.7) and $\tau_{ICC}$, which is used for the continuation analysis.  

We perform a similar procedure for the inner voltage controller, resulting in PI controller gains of
\begin{align}
    &K_{p,IVC} = \frac{6}{\tau_{IVC}\,\omega_b\,X_{c,f}},\\
    &K_{i,IVC} = \frac{9}{\zeta_{IVC}^2\,\tau_{IVC}^2\,\omega_b\,X_{c,f}}.
\end{align}

\subsection{Current Limitation}
\label{sec:mod:limit}
To protect the converter switches from over-current conditions, a circular current limiter can be included \cite{Baeckeland2024_2}. 
This is described as
\begin{equation}
\boldsymbol{i_{lim}^*} = \rho \times \boldsymbol{i^*},
\end{equation}
\begin{equation}
\rho =
\min(1, |\boldsymbol{i}|_{\max} / |\boldsymbol{i^*}|),
\end{equation}
\noindent where $\boldsymbol{i^*}$, $\boldsymbol{i_{lim}^{*}}$, $|\boldsymbol{i}|_{\max}$ are the unsaturated and saturated current references, and the maximum current magnitude, respectively. 
It is also standard practice to include anti-windup control in this scenario to avoid accumulation of the integral output for the inner voltage PI controllers when the current reference is saturated. In particular, we adopt the common back-calculation and tracking method as per \cite{Visioli2006}. This approach requires the tuning of a tracking time constant, $T_s$, for which there exists several different recommendations. We adopt the suggestion from \cite{Visioli2006} where $T_s=K_{i,IVC}$. 

As discussed in Section \ref{sec:meth:bifur}, we investigate the impact of a smooth approximation of the current limiter, as is required for inclusion within continuation analysis \cite{Moutevelis2022}. In particular, we consider the transition from non-limited to limited conditions (and vice versa) as a Heaviside step function approximated based on the hyperbolic tangent function \cite{Moutevelis2022} as
\begin{equation}
    \boldsymbol{i^{*}_{lim}} = (1-\alpha) \times \boldsymbol{i^*} + \alpha\times |\boldsymbol{i}|_{\max}e^{j\angle \boldsymbol{i^*}}
\end{equation}
with
\begin{equation}
    \alpha = \frac{1}{2}\left(1+\tanh\left(\frac{\boldsymbol{\left|i^*\right|}-|\boldsymbol{i}|_{\max}}{2\times \delta}\right)\right)
\end{equation}
where $\delta$ is a ``smoothing'' parameter, the impact of which  
is illustrated in Fig. \ref{fig:LimSmoothApprox}. Particularly, $\delta$ of $0.001$ and $0.01$ provide very close approximations to the ideal (hard) response, while more noticeable, but not ostensibly significant, divergence is seen for $\delta$ of $0.05$ and $0.1$. Please see Appendix \ref{sec:app:currlimit} for time-domain validation of the system response when adopting the current limiter smooth approximation.

\begin{figure}
    \centering
    \includegraphics[width=0.9\linewidth, trim = 47 617 65 30, clip]{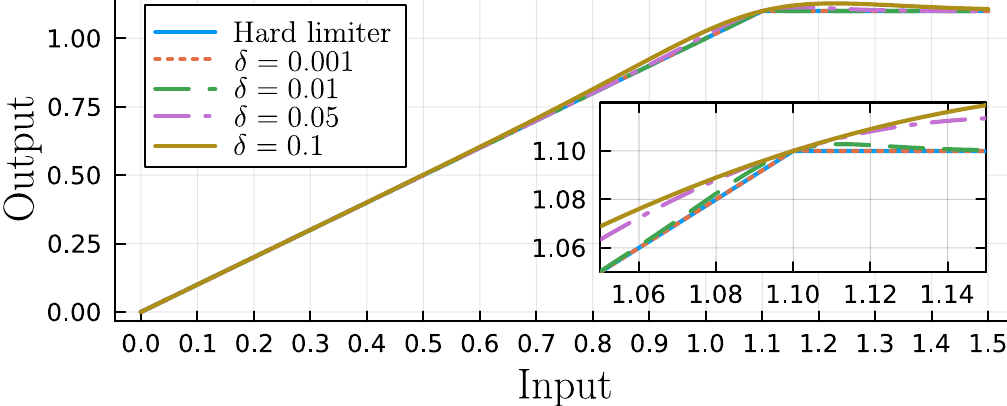}
    \caption{Input-output (unsaturated-saturated current reference in per unit) response of current limiter with different values of the smoothing parameter $\delta$. The inset zooms on the knee point.}
    \label{fig:LimSmoothApprox}
\end{figure}



\section{Bifurcation Analysis and Results}
\label{sec:res}
In this section, parametric stability bounds are illustrated based on continuation analysis of operational parameters and inner controller time constants.
Additionally, we continue the limit cycle associated with the strong grid Hopf bifurcation to reveal rapid onset of unacceptably large oscillations.
Finally, we study the impact of the circular current limiter smooth approximation to show the existence of spurious Hopf bifurcations that can invalidate corresponding analyses.

\subsection{Continuation of Operational Parameters}
\label{sec:res:op}
Beginning with the nominal parameters outlined in Table \ref{tab:params} (current limitation not yet included), we choose the $SCR$ as continuation parameter, and display the equilibrium branch results for differing values of active power set point, $P^*$, in Fig. \ref{fig:BifurDiagram_SCR_Pref}.
\begin{figure}[hbtp]
    \centering
    \begin{subfigure}[b]{ \columnwidth}
		\centering
		\includegraphics[width=\columnwidth,trim=30 595 0 20, clip]{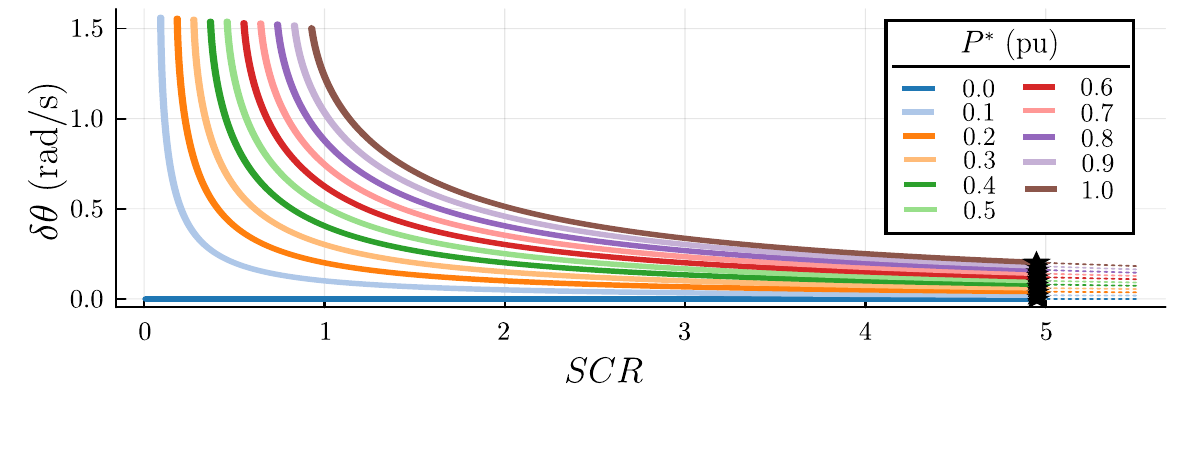}
        \vspace{-10mm}
		\caption{}
	\end{subfigure}
    \begin{subfigure}[b]{ \columnwidth}
		\centering
		\includegraphics[width=\columnwidth,trim=40 605 0 10, clip]{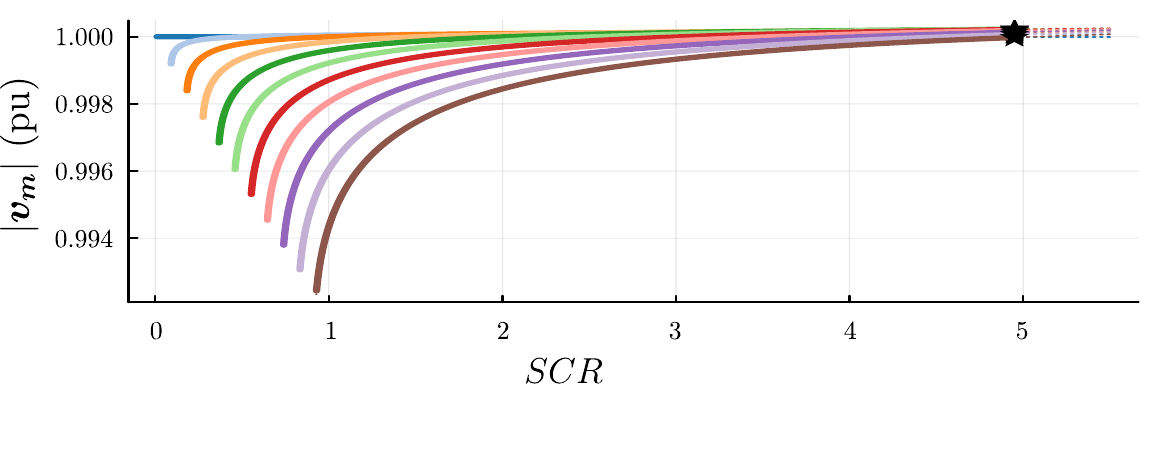}
        \vspace{-10mm}
		\caption{}
	\end{subfigure}
    \caption{Bifurcation diagram for (a) $\left|\boldsymbol{v_m}\right|$ and (b) $\delta\theta$ for varying $SCR$, with different values of $P^*$. A star shows the Hopf point and thin dotted line shows unstable points.}
    \label{fig:BifurDiagram_SCR_Pref}
\end{figure}
Bifurcations are found to occur in both weak grid ($SCR \lesssim 1$) and strong grid scenarios ($SCR \approx 4.95$). The former is the saddle node bifurcation due to the maximum power transfer limits \cite{CutsemVournas} in a lossy system as per \cite{Pepper2024}
\begin{equation}
\label{eq:Pmax}
    P_{\max}=\frac{\left|\boldsymbol{v_m}\right|}{R_g^2+X_g^2}\left( \left|\boldsymbol{v_m}\right|R_g+\left|\boldsymbol{v_g}\right|\sqrt{R_g^2+X_g^2} \right),
\end{equation}
\noindent and with $X_g=10R_g$, $P_{\max}$ occurs at $\delta\theta \approx \angle\boldsymbol{v_m} - \angle\boldsymbol{v_g} = \pi/2 + \tan^{-1}(R_g / X_g)~\text{rad} = 1.6705~\text{rad}$. In this case, as observed in Fig. \ref{fig:BifurDiagram_SCR_Pref}a, the continuation algorithm fails\footnote{Continuation algorithms can fail near the saddle-node point, especially in stiff systems such as the one being studied. This is because the Jacobian becomes increasingly ill-conditioned near the critical point \cite{Ajjarapu}, causing the predictor step to provide poor estimates to the corrector, whose own convergence also degrades due to the ill-conditioning. In this case, extremely small steps of the bifurcation parameter are required to improve the likelihood of Newton convergence, eventually hitting minimum thresholds and computational limits. Furthermore, the fold, as seen in Fig. \ref{fig:BifurDiagram_SCR_Pref}, is very sudden with the voltage magnitude and angle nearing vertical at the failure point.} as we approach the saddle node ($P_{\max}$) point. To confirm this behaviour, we can observe the real part of the rightmost (real-valued) eigenvalue as we approach the weak grid instability point in Fig. \ref{fig:weakgridsaddle_eigs} (for $P^*=0.8~\text{pu}$), where we find that a single real-valued eigenvalue approaches the stability boundary before the continuation corrector step fails to converge. 
Particularly, for a given $X/R$ ratio, $R_g$ and $X_g$ will increase as $SCR$ decreases due to (\ref{eq:SCR}). Therefore, especially for relatively small variations of $\left|\boldsymbol{v_m}\right|$ (Fig. \ref{fig:BifurDiagram_SCR_Pref}b), $P_{\max}$ reduces, and once $P_{\max}$ becomes less than $P^*$, the equilibrium point vanishes. 
\begin{figure}[hbtp]
    \centering
    \includegraphics[width=1.0\linewidth,trim=0 12 0 0, clip]{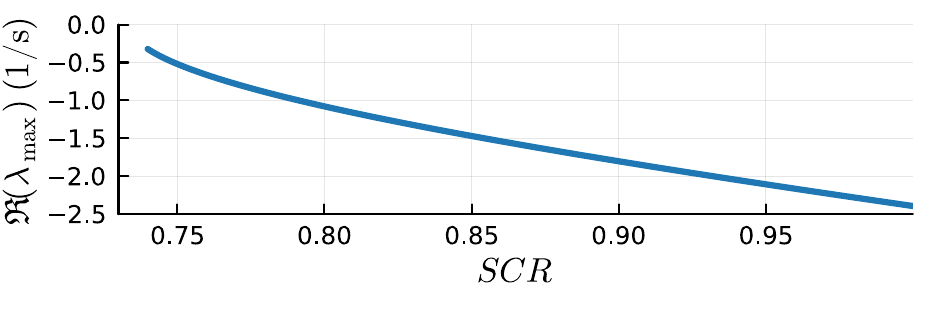}
    \caption{Trajectory of the real part of the rightmost eigenvalue, $\mathfrak{R}(\lambda_{\max})$, as the $SCR$ is varied, with $P^*=0.8~\text{pu}$.}
    \label{fig:weakgridsaddle_eigs}
\end{figure}

The strong grid instability is related to the Hopf bifurcation associated with the inner controller-related SSO\footnote{The inner controller characteristics of these SSOs are discussed in \cite{Sharjeel2025_2} and \cite{Benedetti2025} and, although not displayed for brevity, confirmed in this work through participation factor analysis.}, and is seen to be largely unaffected by $P^*$. 
Considering converter-interfaced generators may be connected at different voltage levels, with different $X/R$ ratios, we also perform codim-2 continuation (Section \ref{sec:meth:cont}) of this strong grid Hopf point, varying the $X/R$ ratio along with the $SCR$. This is displayed in Fig. \ref{fig:BifurDiagram_SCR_XR}, with the colourbar relaying the frequency of the SSO along the Hopf branch. 
We find that decreasing the $X/R$ ratio increases the $SCR$ at which the strong grid instability occurs, suggesting that distribution-connected GFMs utilising the prototypical control scheme of Fig. \ref{fig:GFMcontrol} can be connected in stronger grids than transmission system-connected GFMs.
We can also note that the frequency of the corresponding SSO reduces with the reduction of $X/R$ ratio (and concomitant increase of $SCR$). At the nominal $X/R = 10$, the frequency is $25.58~\text{Hz}$, whereas this reduces to $5.71~\text{Hz}$ at $X/R = 0.1$.
\begin{figure}[hbtp]
    \centering
    \includegraphics[width=\columnwidth,trim=5 640 17.5 10, clip]{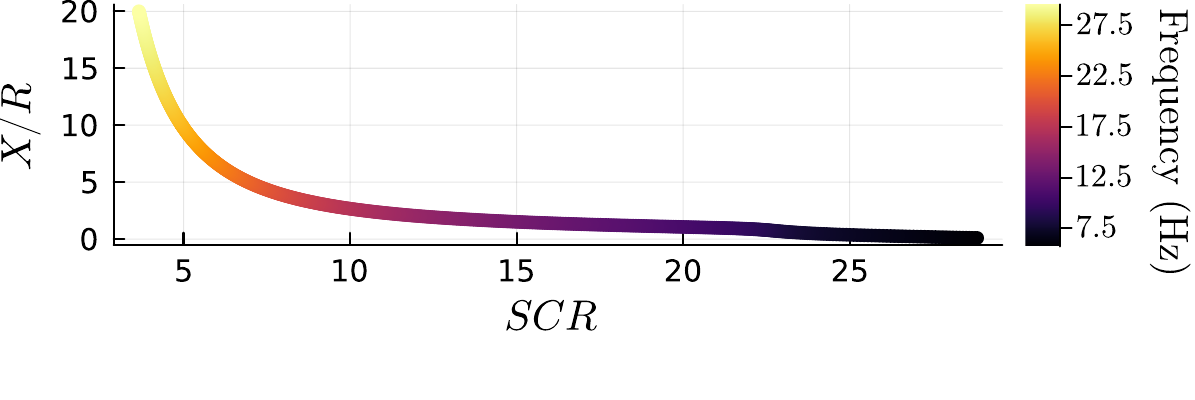}
    \vspace{-10mm}
    \caption{Bifurcation diagram for simultaneous variation of $SCR$ and $X/R$ ratio, following the strong grid Hopf bifurcation. 
    }
    \label{fig:BifurDiagram_SCR_XR}
\end{figure}

\subsection{Continuation of the Limit Cycle}
Returning to the nominal parameters of $X/R=10$ and $P^*=0.8~\text{pu}$, we can apply the normal form analysis (Section \ref{sec:meth:contPO}) to the strong grid Hopf bifurcation. We find that the first Lyapunov coefficient, denoted as $b$ in (\ref{eq:HopfNormalForm}), is equal to $-0.00198-j0.00096$. Since the real part of $b$ is negative, the Hopf bifurcation is supercritical\footnote{The supercriticality of the Hopf bifurcation is consistent across the different parameter conditions analysed in this work, except when adopting the smooth current limiter approximation, as per Appendix \ref{sec:app:spur}.} and a stable limit cycle emerges as the parameter continues past the bifurcation point.

We can then perform continuation of the limit cycle (Section \ref{sec:meth:contPO}) and observe the impact on $\left| \boldsymbol{v_m} \right|$ and $\left| \boldsymbol{i_{cv}} \right|$. The corresponding maximum and minimum of the magnitude of the limit cycle are displayed in Fig. \ref{fig:PO_cont}, revealing that very small increase of the $SCR$ past the bifurcation point results in rapid growth of the limit cycle, quickly reaching unacceptable levels. 
Note, $|\boldsymbol{i_{cv}}|$ constitutes a magnitude which is why the lower bound of the limit cycle appears to ``bounce'' off of zero, when it is rather changing direction.

\begin{figure}[hbtp]
    \centering
        \includegraphics[width=1.0\linewidth,trim=5 650 15 20, clip]{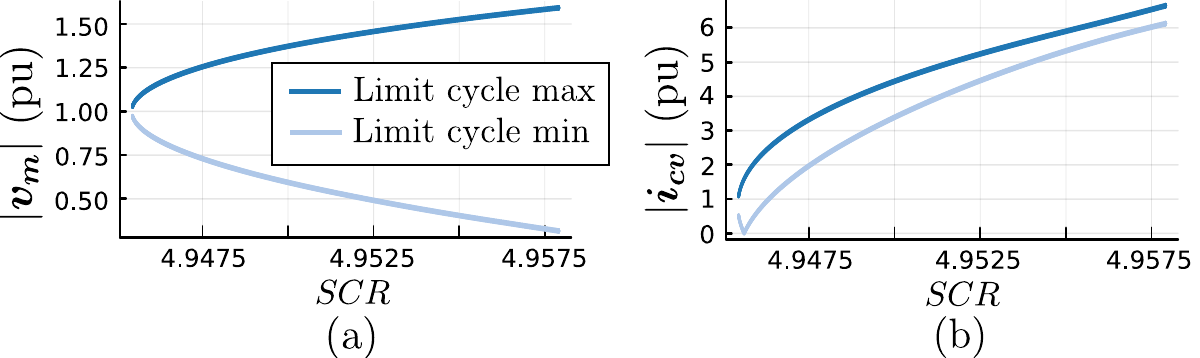}
    \caption{Bifurcation diagram showing the maximum and minimum bounds of the limit cycle as $SCR$ is continued past the strong grid Hopf bifurcation point.}
    \label{fig:PO_cont}
\end{figure}

\subsection{Continuation of Inner Controller Time Constants}
As previously discussed, the strong grid instability-inducing SSO results from an interaction involving the cascaded inner voltage and current controllers \cite{Sharjeel2025_2, Benedetti2025}. Therefore, in this subsection we perform codim-2 continuation analysis of the Hopf bifurcation with respect to the inner controller time constants $\tau_{IVC}$ and $\tau_{ICC}$ (Section \ref{sec:mod:inner}). The continuation plot, with the nominal parameters of Table \ref{tab:params}, is displayed in 
in Fig. \ref{fig:innerBW}. 
Clearly, the parametric stability bounds indicate that the time constants for the inner voltage and current controllers must be sufficiently large and small, respectively, highlighting the need for sufficient timescale separation in strong grid-connected GFM control designs. Consequently, if the bandwidth of the inner current controller is limited due to reduced switching frequency of the converter, the inner voltage controller must be sufficiently slow to maintain stability. However, if $\tau_{ICC}\gtrapprox 2.5~\text{ms}$ when $SCR=5$, then tuning $\tau_{IVC}$ cannot bring stability in strong grids.
\begin{figure}[hbtp]
    \centering
    \includegraphics[width=0.9\linewidth,trim=8 705 150 10, clip]{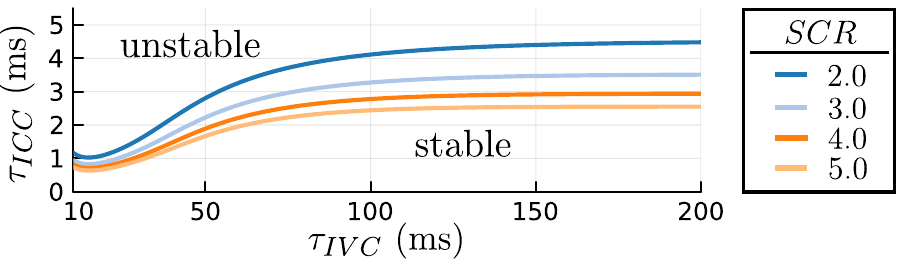}
    \caption{Bifurcation diagram for simultaneous variation of $\tau_{IVC}$ and $\tau_{ICC}$ ratio, following the strong grid Hopf bifurcation.}
    \label{fig:innerBW}
\end{figure}

\subsection{Impact of Current Limiter Smooth Approximation}
\label{sec:res:lim}
In this subsection, we re-perform the $SCR$ continuation analysis from Section \ref{sec:res:op} but with the addition of the circular current limiter outlined in Section \ref{sec:mod:limit}. We consider both the ideal (hard) limiter and the ostensibly very good smooth approximations of $\delta=0.001$ and $\delta=0.01$, with $P^*=1~\text{pu}$. The continuation results for the smooth approximations are displayed in Fig. \ref{fig:CurrLimitSmooth}, while the hard limiter continuation 
gives the same curve shape as the $\delta=0.001$ but without the Hopf points (included in Appendix \ref{sec:app:spur} for completeness), and it fails at the limit-induced bifurcation point. In particular, this spurious Hopf bifurcation occurs in the weak grid scenario when $SCR$ is reduced to $1.043$ and $1.234$ for $\delta=0.001$ and $\delta=0.01$ respectively, despite not occurring when the ideal (hard) limiter is modelled. The corresponding complex conjugate pair of eigenvalues then returns to stability just before the limit-induced bifurcation point \cite{Xing2021} (at $I_{cv}=1.1~\text{pu}$ and $SCR\approx1.033$) when $SCR=1.034$ and $SCR=1.048$ for $\delta=0.001$ and $\delta=0.01$ respectively. 
Furthermore, analysing the corresponding eigenvalue trajectory in Fig. \ref{fig:smoothEigs} reveals that this Hopf bifurcation is induced by the same SSO that causes the strong grid Hopf bifurcation. 
It can also be noted that the strong grid Hopf bifurcation appears at a lower value of $SCR = 4.390$ when $\delta=0.01$.
\begin{figure}[hbtp]
    \centering
    \begin{subfigure}[b]{ \columnwidth}
		\centering
		\includegraphics[width=0.9\columnwidth,trim=85 695 72 20, clip]{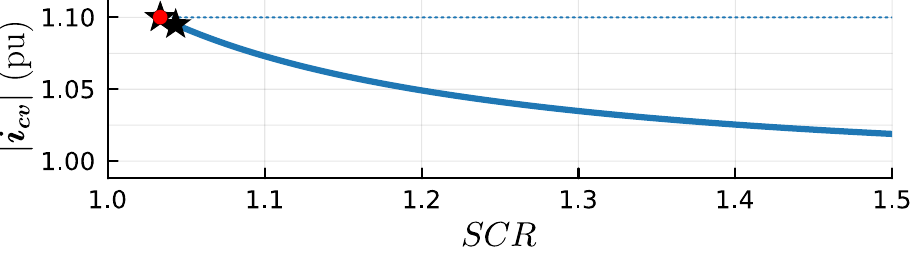}
		\caption{}
	\end{subfigure}
    \begin{subfigure}[b]{ \columnwidth}
		\centering
		\includegraphics[width=0.9\columnwidth,trim=85 695 72 20, clip]{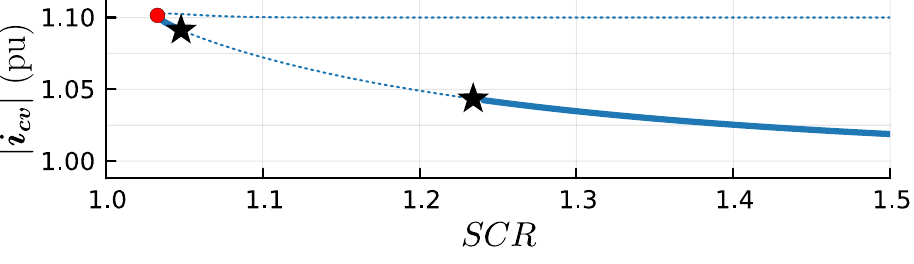}
		\caption{}
	\end{subfigure}
    
    \caption{Bifurcation diagram of $\left| \boldsymbol{i_{cv}} \right|$ against $SCR$ with smooth limiter with (a) $\delta=0.001$ and (b) $\delta=0.01$. Black stars and red circles show Hopf and limit-induced bifurcation points, respectively. Thin dotted lines denote instability.}
    \label{fig:CurrLimitSmooth}
\end{figure}
\begin{figure}[hbtp]
    \centering
    \includegraphics[width=0.9\linewidth, trim = 47 660 37 25, clip]{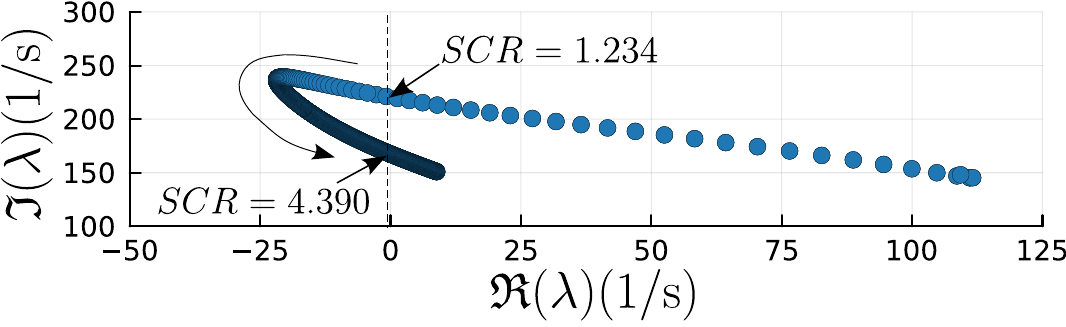}
    \caption{Trajectory of the strong grid instability-inducing SSO eigenvalue, $\lambda$, as $SCR$ varies between $1.075$ and $5.5$ with circular current limiter smooth approximation with $\delta=0.01$.}
    \label{fig:smoothEigs}
\end{figure}

Further analysis of this spurious Hopf bifurcation can be found in Appendix \ref{sec:app:spur}, also revealing that the smooth approximation can change the strong grid (and corresponding weak grid) Hopf to subcritical.
However, this ultimately constitutes a modelling artefact (i.e., it is not present when using the hard limiter model) that can invalidate the results of bifurcation analysis, suggesting false instabilities. Subsequently, extreme caution, and careful validation is recommended if adopting smooth current limiter approximations.  


\section{Conclusions}
\label{sec:conc}
This paper details a continuation-based bifurcation analysis of a grid-tied multi-loop controlled GFM converter. In particular, focus is given to an inner controller-related SSO that can incite instability in strong grids. 
Parametric stability bounds are outlined for a range of $SCR$, $X/R$ ratio, and active power set point conditions, as well as highlighting the need for sufficiently small and large time constants for the inner current and voltage controllers, respectively.

Additionally, we reveal through normal form theory that the corresponding Hopf bifurcation is supercritical, and find that subsequent continuation of the bifurcation parameter leads to rapid onset of unacceptably large limit cycle oscillations (e.g., extensive voltage and current swings).

Finally, we find that the adoption of smooth approximations of the (circular) current limiter can cause spurious Hopf bifurcations of the aforementioned SSO in weak(er) grid conditions, representing a modelling artefact, even for ostensibly sufficient approximations of the ideal saturation response.

\appendix
\subsection{Time-Domain Validation of the Circular Current Limiter Smooth Approximation}
\label{sec:app:currlimit}
To add to the static input-output visualisation of Fig. \ref{fig:LimSmoothApprox}, we consider here the time-domain response of the system under test in response to a grid voltage magnitude dip from $1~\text{pu}$ to $0.9~\text{pu}$ which starts at $0.01~\text{s}$ and lasts $20~\text{ms}$. Parameters are set to nominal values of Table \ref{tab:params} but with $P^*=1~\text{pu}$. It can be seen from Fig. \ref{fig:TimeDomainVal} that $\delta=0.001$ provides a very good approximation to the ideal hard current limiter response, while $\delta=0.01$ results in some slight but not significant discrepancies.
When adopting $\delta$ of $0.05$ or $0.1$, the time-domain response shows instability, the reason for which can be understood based on the spurious Hopf bifurcations as discussed in Section \ref{sec:res:lim} and Appendix \ref{sec:app:spur}. 

\begin{figure}[hbtp]
    \centering
    \includegraphics[width=1.0\linewidth, trim = 75 670 80 20, clip]{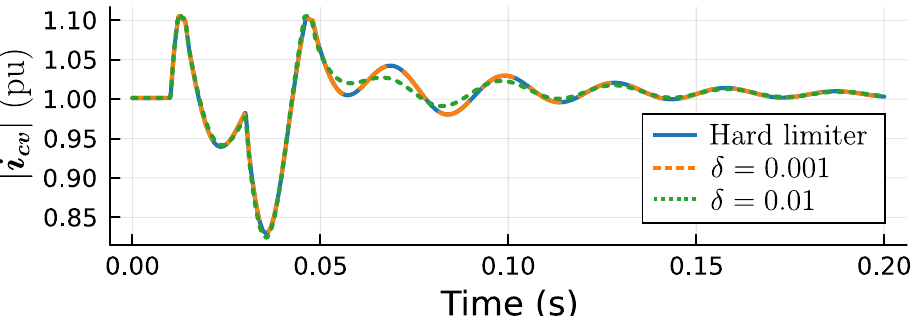}
    \caption{Time-domain response of $\left| \boldsymbol{i_{cv}} \right|$ when the system is subject to a grid voltage dip, when including current limiter with ideal (hard) response as well as smooth approximation with $\delta=0.001$ and $\delta=0.01$.}
    \label{fig:TimeDomainVal}
\end{figure}




\subsection{Supplementary Analysis of the Spurious Hopf Bifurcation}
\label{sec:app:spur}
For completeness, Fig. \ref{fig:hardLimCont} displays the SCR continuation with hard limiter model, for comparison against Fig. \ref{fig:CurrLimitSmooth}. This confirms the lack of weak grid Hopf bifurcation, and the failure of the continuation at the limit-induced bifurcation point (i.e., no subsequent thin dotted line as in the smooth approximation case).
\begin{figure}[hbtp]
    \centering
    \includegraphics[width=0.9\columnwidth,trim=85 695 72 20, clip]{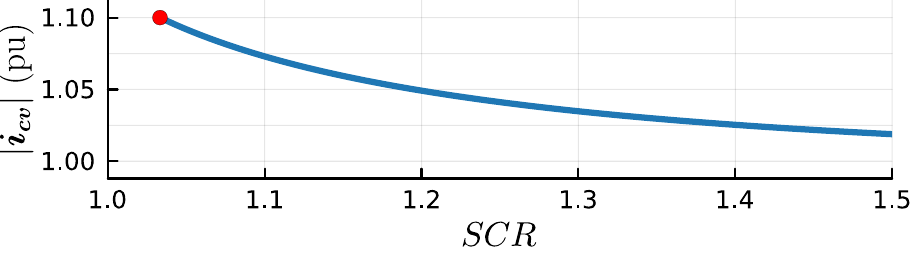}
    \caption{Bifurcation diagram of $\left| \boldsymbol{i_{cv}} \right|$ against $SCR$ with hard current limiter model. The red circle shows the limit-induced bifurcation point.}
    \label{fig:hardLimCont}
\end{figure}

We can continue the (spurious) weak grid Hopf point by varying the smoothing parameter $\delta$ along with the $SCR$ as per Fig. \ref{fig:delta_cont}. This reveals that further increase of $\delta$ increases (worsens) the lower $SCR$ stability boundary associated with this Hopf bifurcation. At the same time, we observe that the weak and strong grid Hopf bifurcations approach each other as $\delta$ is increased, until they meet at $\delta = 0.0142$ and any further increase results in instability at all points for the range of $SCR$ considered. This behaviour can be intuitively (yet crudely) understood by considering the whole eigenvalue curve of Fig. \ref{fig:smoothEigs} moving to the right as we increase $\delta$.

It can also be noted that the strong grid Hopf bifurcation remains supercritical only for very small values of $\delta$ with a Bautin (generalized Hopf) bifurcation occurring when $\delta\approx 0.0011$ and $SCR\approx 4.944$, after which (i.e., smaller $SCR$) the bifurcation is instead subcritical. The (spurious) weak grid Hopf bifurcation is always subcritical.

\begin{figure}[hbtp]
    \centering
    \includegraphics[width=.9\linewidth, trim=60 625 90 35, clip]{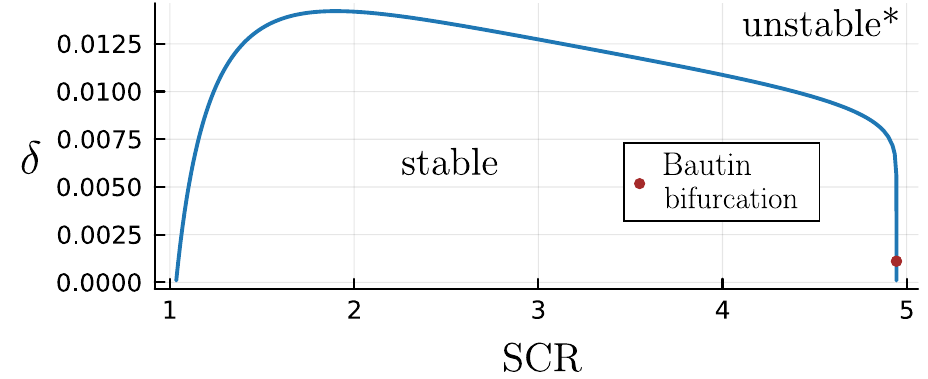}
    \caption{Bifurcation diagram for simultaneous variation of $SCR$ and $\delta$, following the weak grid Hopf bifurcation (which is seen to lead into the strong grid Hopf) when adopting smooth approximation of the circular current limiter. Note, the SSO returning to stability (at $SCR=1.043$) just before the limit-induced bifurcation point, as discussed in Section \ref{sec:res:lim}, means that the stability of the system is not solely delineated by the Hopf curve, hence the * superscript in the unstable label.}
    \label{fig:delta_cont}
\end{figure}

\bibliographystyle{IEEEtran}
\bibliography{bibFile}
\balance

\endgroup
\end{document}